\magnification \magstep1
\raggedbottom
\openup 4\jot
\voffset6truemm
\headline={\ifnum\pageno=1\hfill\else
\hfill {\it 1-loop effective action on the 4-ball} 
\hfill \fi}
\def\cstok#1{\leavevmode\thinspace\hbox{\vrule\vtop{\vbox{\hrule\kern1pt
\hbox{\vphantom{\tt/}\thinspace{\tt#1}\thinspace}}
\kern1pt\hrule}\vrule}\thinspace}
\centerline {\bf 1-LOOP EFFECTIVE ACTION ON THE 4-BALL}
\vskip 0.3cm
\centerline {Giampiero Esposito$^{1,2}$,
Alexander Yu Kamenshchik$^{3}$
and Giuseppe Pollifrone$^{4}$}
\vskip 0.3cm
\centerline {\it ${ }^{1}$Istituto Nazionale di Fisica Nucleare,
Sezione di Napoli,}
\centerline {\it Mostra d'Oltremare Padiglione 20,
80125 Napoli, Italy}
\centerline {\it ${ }^{2}$Dipartimento di Scienze Fisiche,}
\centerline {\it Mostra d'Oltremare Padiglione 19,
80125 Napoli, Italy}
\centerline {\it ${ }^{3}$Nuclear Safety Institute, Russian
Academy of Sciences,}
\centerline {\it Bolshaya Tulskaya 52, Moscow 113191, Russia}
\centerline {\it ${ }^{4}$Dipartimento di Fisica, Universit\`a
di Roma ``La Sapienza",}
\centerline {\it and INFN, Sezione di Roma, Piazzale Aldo
Moro 2, 00185 Roma, Italy}
\vskip 0.3cm
\noindent
{\bf Abstract.} This paper applies $\zeta$-function regularization 
to evaluate the 1-loop effective action for 
scalar field theories and Euclidean Maxwell
theory in the presence of boundaries. After a comparison of
two techniques developed in the recent literature, vacuum 
Maxwell theory is studied and the contribution of all
perturbative modes to $\zeta'(0)$
is derived: transverse, longitudinal and
normal modes of the electromagnetic potential, jointly with
ghost modes. The analysis is performed on imposing magnetic 
boundary conditions, when the Faddeev-Popov Euclidean 
action contains the particular gauge-averaging term which leads
to a complete decoupling of all perturbative modes. It is
shown that there is no cancellation of the contributions to
$\zeta'(0)$ resulting from longitudinal, normal and ghost modes.
\vskip 100cm
\leftline {\bf 1. Introduction}
\vskip 0.3cm
\noindent
The approach to quantum field theory and quantum gravity in terms
of the effective action has led to many deep insights into the
structure of physical theories [1--8]. Over the last few years,
a number of papers appeared in which the 1-loop effective action
was calculated on non-trivial background geometries, including
those with boundaries [7--12]. Here we carry out calculations of
such a kind on the 4-ball for scalar fields and for Euclidean 
Maxwell theory, taking into account in the last case also the
contributions of ghost and gauge modes.

Now we would like to recall the basic definitions concerning
the effective action. For simplicity, we describe
in this section the basic equations and ideas in the absence of
gauge groups. The starting point is a functional-integral
representation of the $<{\rm out} \mid {\rm in}>$ amplitude
as [5]
$$
<{\rm out} \mid {\rm in}>=e^{iW[J]}
=N \int e^{i(S[\varphi]+J_{l}\varphi^{l})}
\mu[\varphi]d\varphi
\eqno (1.1)
$$
where the integral is over all superclassical fields satisfying
the given boundary conditions, $S[\varphi]$ is the classical
action supplemented by boundary terms appropriate to the 
{\it in} and {\it out} eigenvectors, $J_{i}$ are external
sources coupled to the field operators, and $\mu[\varphi]$
is a suitable measure for integration [5]. One may then define
the effective action as the Legendre transform of $W[J]$ as
$$
\Gamma[{\overline \varphi}] 
\equiv W[J] -J_{l}{\overline \varphi}^{l} .
\eqno (1.2)
$$
This action obeys the functional equation [5]
$$
\Gamma_{,i}[{\overline \varphi}]=-J_{i}
\eqno (1.3)
$$
and an equivalent formulation of the theory may be based
on the two equations [5]
$$
e^{i\Gamma[{\overline \varphi}]}
=N \int 
{\exp} \; i \left \{S[\varphi]
+\Gamma_{,l}[{\overline \varphi}]
\Bigr({\overline \varphi}^{l}-\varphi^{l}\Bigr)
\right \} \; \mu[\varphi] \; d\varphi
\eqno (1.4)
$$
$$
<A[\varphi]>=Ne^{-i\Gamma[{\overline \varphi}]}
\int A[\varphi]  
{\exp} \; i \left \{S[\varphi]
+\Gamma_{,l}[{\overline \varphi}]
\Bigr({\overline \varphi}^{l}-\varphi^{l}\Bigr)
\right \} \; \mu[\varphi] \; d\varphi
\eqno (1.5)
$$
where $<A[\varphi]>$ is the standard notation for
chronological averages [3, 5]. Everything in the theory 
can indeed be derived from (1.4) and (1.5), provided that
${_{i,}\Gamma_{,j}}[{\overline \varphi}]$ is a non-singular
integro-differential operator, which ensures that 
${\overline \varphi}^{i}$ coincides with the chronological
average of ${\varphi}^{i}$. Moreover, by paying attention
to the measure, the chronological average of the operator
field equations is found to be [5]
$$
<S_{,j}[\varphi]-i(\log \mu[\varphi])_{,j}>
-\Gamma_{,j}[{\overline \varphi}]=0 .
\eqno (1.6)
$$
The extension to gauge fields and gravitation may be
found, for example, in [3--6], and it leads to a powerful
and elegant formulation of perturbative quantum field theory.

In recent years, as already mentioned above, 
motivated by the analysis of semi-classical
effects in quantum field theory, impressive progress has been
made in the calculation of the 1-loop effective action [7--12].
In the geometric approach [7], the 1-loop approximation is
obtained after a careful application of the Schwinger-DeWitt
method [1, 2, 13], and the information on the non-local part of the
1-loop effective action is encoded in the form factors
(see [7, 8] and references therein). In the analytic approach,
which has been successful in the more difficult case of 
manifolds with boundary, one applies $\zeta$-function regularization,
the uniform asymptotic expansions of Bessel functions 
for the flat-space case, or Legendre functions for
the 4-sphere case, and
suitable contour formulae to evaluate functional determinants
for scalar, spinor and gauge fields [10--12].

For a given elliptic operator $\cal A$, the corresponding
zeta-function $\zeta_{\cal A}$ is defined as the trace of its
complex power ${\cal A}^{-s}$, and admits an analytic
continuation to the whole complex-$s$ plane as a meromorphic
function [14]. 
It is then possible to make sense of $\zeta_{\cal A}(0)$
and $\zeta_{\cal A}'(0)$, and the 1-loop effective action 
in four dimensions takes the form [7, 8]
$$
\Gamma_{\cal A}^{(1)}=-{1\over 2}\zeta_{\cal A}'(0)
-{1\over 2}\zeta_{\cal A}(0)\log(\mu^{2}) .
\eqno (1.7)
$$
In physical applications, $\cal A$ is the Laplace operator, or the
squared Dirac operator, or a matrix of elliptic operators in the
case of gauge fields and gravitation, where gauge and ghost modes
also contribute to the full 1-loop effective action.

In the analysis of manifolds with boundary, complete results for
$\zeta'(0)$ have been obtained in [9--12] for scalar and spin-1/2
fields, but for gauge fields
and gravitation only the contributions of physical degrees 
of freedom (i.e. transverse or transverse-traceless modes) 
were obtained in [11, 12, 15--17]. The aim of our paper has been
therefore to complete the recent investigations of the 1-loop
effective action for gauge fields in the presence of boundaries.
Since the calculations are, in general, extremely lengthy (at
least if one wants to double-check them by hand, which remains
a necessary step), we have focused on the simplest (but
non-trivial) model, i.e. Euclidean Maxwell theory in the presence
of boundaries [18--21]. Attention is focused on the particular 
gauge-averaging functional studied in [21], since this was found
to lead to a trace anomaly which agrees with the one resulting from
the transverse modes only [22], on imposing magnetic boundary
conditions. It was then necessary to understand whether the 
cancellation of the contributions of gauge and ghost modes in
this gauge is also a property of the full 1-loop effective
action.

For this purpose, section 2 describes the application of the 
algorithm by Bordag et al. [10] to four-dimensional Euclidean
backgrounds with boundary. Section 3 studies instead the
Barvinsky-Kamenshchik-Karmazin-Mishakov technique [15--17] 
in the case of Dirichlet boundary conditions.
Section 4 obtains the 1-loop
effective action with magnetic boundary conditions. 
As we said before, the gauge
chosen is the one first derived in [21], where all perturbative
modes for the electromagnetic potential and its ghost are
decoupled. Concluding remarks and open 
problems are presented in section 5, and relevant details
are given in the appendix.
\vskip 0.3cm
\leftline {\bf 2. Bordag-Geyer-Kirsten-Elizalde algorithm}
\vskip 0.3cm
\noindent
A powerful analytic algorithm for the
calculation of $\zeta'(0)$ for manifolds with boundary
is the one developed by Bordag et al. in [10]. We here
describe its application to the analysis of the Laplace
operator acting on massless scalar fields in flat
four-dimensional Euclidean space bounded by a 3-sphere of
radius $a$, in the case of Dirichlet boundary conditions. 
This is indeed the background considered in the original
calculations of the following sections, and in [18--21].

The work of [10] shows that the $\zeta$-function of such a
four-dimensional boundary-value problem may be expressed as
the sum of two terms as
$$
\zeta(s)=\sum_{l=0}^{\infty}(l+1)^{2}Z_{l+1}(s)
+\sum_{i=-1}^{3}A_{i}(s)
\eqno (2.1)
$$
where, with the notation in the appendix, one has
$$ \eqalignno{
Z_{l+1}(s)&={\sin(\pi s)\over \pi}\int_{0}^{\infty}dz
{\left({z(l+1)\over a}\right)}^{-2s}{\partial \over
\partial z} \biggr[\log I_{l+1}((l+1)z)-(l+1)\eta \cr
&+\log \Bigr(\sqrt{2\pi (l+1)}(1+z^{2})^{1\over 4}\Bigr)
-\sum_{n=1}^{3}{D_{n}(t)\over (l+1)^{n}} \biggr]
&(2.2)\cr}
$$
$$
A_{-1}(s)={1\over 4 \sqrt{\pi}} {a^{2s}\over \Gamma(s)}
{\Gamma(s-{1\over 2})\over s}
\zeta_{R}(2s-3)
\eqno (2.3)
$$
$$
A_{0}(s)=-{1\over 4} a^{2s}\zeta_{R}(2s-2)
\eqno (2.4)
$$
$$
A_{1}(s)=-{1\over 2} {a^{2s}\over \Gamma(s)}\zeta_{R}(2s-1)
\sum_{j=0}^{1}x_{1,j}(1+2j)
{\Gamma(s+j+{1\over 2})\over \Gamma(j+{3\over 2})}
\eqno (2.5)
$$
$$
A_{2}(s)=-{1\over 2} {a^{2s}\over \Gamma(s)}\zeta_{R}(2s)
\sum_{j=0}^{2}x_{2,j}(2+2j)
{\Gamma(s+j+1)\over \Gamma(j+2)}
\eqno (2.6)
$$
$$
A_{3}(s)=-{1\over 2} {a^{2s}\over \Gamma(s)}\zeta_{R}(2s+1)
\sum_{j=0}^{3}x_{3,j}(3+2j)
{\Gamma(s+j+{3\over 2})\over \Gamma (j+{5\over 2})} .
\eqno (2.7)
$$
The $\zeta'(0)$ value is then obtained by differentiation
at $s=0$ of (2.1), making use of (2.2)--(2.7), 
of the identity [10]
$$ \eqalignno{
Z_{l+1}'(0)&=\log(\Gamma(l+2))+(l+1)-(l+1)\log(l+1) \cr
&-{1\over 2}\log(2\pi(l+1))
+\sum_{n=1}^{3}{D_{n}(1)\over (l+1)^{n}} \cr
&=\int_{0}^{\infty} dt \biggr({1\over 2}-{1\over t}
+{1\over (e^{t}-1)}\biggr){e^{-t(l+1)}\over t}
+\sum_{n=1}^{3}{D_{n}(1)\over (l+1)^{n}}
&(2.8)\cr}
$$
and of the equations (A.4)--(A.6) of the appendix. The resulting
$\zeta'(0)$ value is found to be [10]
$$
\zeta'(0)={173\over 30240}+{1\over 90}\log(2/a)
+{1\over 3}\zeta_{R}'(-3)
-{1\over 2}\zeta_{R}'(-2)
+{1\over 6}\zeta_{R}'(-1).
\eqno (2.9)
$$
\vskip 0.3cm
\leftline {\bf 3. Barvinsky-Kamenshchik-Karmazin-Mishakov
technique}
\vskip 0.3cm
\noindent
With the notation described in detail in [15--17], and applied
also in [19], the $\zeta'(0)$ value in field theory takes the
form
$$
\zeta'(0)=I_{\rm reg}^{R}(M=\infty)-I^{R}(M=0)
-\int_{0}^{\infty}dM^{2} \; \log(M^{2})
{dI_{\rm pole}(M^{2})\over dM^{2}}.
\eqno (3.1)
$$
In the case of a real, massless scalar field subject to
homogeneous Dirichlet conditions on the 3-sphere, the 
BKKM function of [15--17] reads
$$
I(M^{2},s) \equiv \sum_{n=1}^{\infty}n^{2-2s}
\log f_{n}(nMa)
\eqno (3.2)
$$
where $f_{n}(nMa)=(Ma)^{-n}I_{n}(nMa)$. Thus, with the 
notation of equation (A.2) of the appendix, one finds
$$ \eqalignno{
I_{\rm pole}&={1\over 6}u_{1}^{3}(t)+{1\over 2}u_{3}(t)
-{1\over 2}u_{1}(t)u_{2}(t) \cr
&={(5525t^{9}-9945t^{7}+4779t^{5}-375t^{3})\over
11520}
&(3.3)\cr}
$$
where $t \equiv 1/\sqrt{1+z^{2}a^{2}}$.
It is now useful to set $x \equiv M^{2}a^{2}$, and then
re-express the integral in (3.1) as
$$
{\cal A}=\log(a^{2}) \Bigr(I_{\rm pole}(\infty)-I_{\rm pole}(0)\Bigr)
-\int_{0}^{\infty}{dI_{\rm pole}\over dx}\log(x) \; dx.
\eqno (3.4)
$$
The integral in (3.4) can be evaluated with the help of
equations (A.27)--(A.31) of the appendix. These lead to
$$
{\cal A}=\log(a^{2}) \Bigr(I_{\rm pole}(\infty)-I_{\rm pole}(0)\Bigr)
+{1\over 360}\log(2)+{47\over 30240}.
\eqno (3.5)
$$

Moreover, by virtue of the uniform asymptotic expansion (A.2),
one finds
$$
I_{\rm reg}^{R}(M=\infty)=I_{\rm log} \log(a^{2})
-{1\over 2}\log(2\pi) \zeta_{R}(-2).
\eqno (3.6)
$$

The calculation of $I^{R}(0)$ makes it necessary to use
the representation of Bessel functions in terms of
hypergeometric functions, which implies
$$
I^{R}(0)=-\sum_{n=1}^{\infty}n^{2}\Bigr(n\log(2)
+\log \Gamma(n+1) \Bigr).
\eqno (3.7)
$$
Thus, Stirling's formula for the logarithm of the $\Gamma$
function (cf (A.7)), jointly with the careful treatment
of the limit as $s \rightarrow 0$ resulting from (3.2),
lead to [23]
$$ \eqalignno{
I_{\rm reg}^{R}(M=\infty)-I^{R}(0)&=
I_{\rm log} \log(a^{2})+{1\over 120}\log(2)
-\zeta_{R}'(-3)-{1\over 2}\zeta_{R}'(-2)
-{1\over 120} \cr
&+{1\over 2}\sum_{n=1}^{\infty}n^{2-2s}\sum_{k=1}^{\infty}
{B_{2k}\over (2k-1)k \; n^{2k-1}}.
&(3.8)\cr}
$$
In the double sum in (3.8), only the terms with $k=1$
and $2$ have divergent behaviour, and should be treated
separately. They contribute
$-{1\over 144}-{1\over 360}{1\over s}
+{1\over 180}\gamma$,
where $-{1\over 360}{1\over s}$ should be excluded because it
belongs to $I_{\rm pole}(M^{2})$. Moreover, the double infinite 
sum from $k=3$ to $\infty$ in (3.8) contributes 
${4\over 3}\zeta_{R}'(-3)+{1\over 6}\zeta_{R}'(-1)
+{7\over 360}-{1\over 180}\gamma$.
Hence one finds (cf [10])
$$ \eqalignno{
I_{\rm reg}^{R}(M=\infty)-I^{R}(0)&=I_{\rm log} \log(a^{2})
+{1\over 120}\log(2) \cr
&+{1\over 3}\zeta_{R}'(-3)
-{1\over 2}\zeta_{R}'(-2)
+{1\over 6}\zeta_{R}'(-1)+{1\over 240}
&(3.9)\cr}
$$
and the $\zeta'(0)$ value is finally obtained from (3.5)
and (3.9) as
$$
\zeta'(0)={173\over 30240}+
\zeta(0)\log(a^{2}/4)+{1\over 3}\zeta_{R}'(-3)
-{1\over 2}\zeta_{R}'(-2)
+{1\over 6}\zeta_{R}'(-1)
\eqno (3.10)
$$
bearing in mind that [15--17] $\zeta(0)=I_{\rm log}
+I_{\rm pole}(\infty)-I_{\rm pole}(0)$, which equals
$-{1\over 180}$ in our case. 
\vskip 0.3cm
\leftline {\bf 4. Magnetic boundary conditions for
Euclidean Maxwell theory}
\vskip 0.3cm
\noindent
In this section we evaluate the 1-loop 
effective action for Euclidean Maxwell theory on the flat 
four-dimensional background bounded by a 3-sphere, 
on using the particular gauge averaging 
functional described in [21], which enables one to 
decouple all the modes from the beginning. 

For this purpose, one expands the 
normal and tangential components of 
the electromagnetic potential on a family of 3-spheres as
$$
A_{0}(x,\tau)=\sum_{n=1}^{\infty}R_{n}(\tau)Q^{(n)}(x) 
\eqno (4.1)
$$
$$
A_{k}(x,\tau)=\sum_{n=2}^{\infty}\Bigr[f_{n}(\tau)S_{k}^{(n)}(x)
+g_{n}(\tau)P_{k}^{(n)}(x)\Bigr]
\; \; \; \; {\rm for} \; {\rm all} \; k=1,2,3 
\eqno (4.2)
$$
where $Q^{(n)}(x),S_{k}^{(n)}(x),P_{k}^{(n)}(x)$ are 
scalar, transverse and longitudinal vector harmonics on $S^3$, 
respectively [24]. Within the framework of Faddeev-Popov formalism, 
on choosing the gauge-averaging functional as [21] 
$$
\Phi_{P}(A)\equiv {}^{(4)}\nabla^{\mu} A_{\mu} - {2 \over 3} 
A_{0} {\rm Tr} K 
\eqno (4.3)
$$
and by setting to 1 the parameter $\alpha$ occurring in 
the total Euclidean action [18], one eventually 
gets the regular 
solutions for normal and longitudinal components of 
the electromagnetic potential and the 
basis functions for the ghost (denoted by $\epsilon_{n}(\tau)$)
$$
g_{n}(\tau)=A I_{\nu}(M\tau) 
\eqno (4.4)
$$
$$
R_{n}(\tau)=B{1 \over \tau}I_{\nu}(M\tau)  
\eqno (4.5)
$$
$$
\epsilon_{n}(M\tau) = C I_{\nu}(M\tau)  
\eqno (4.6)  
$$
where $\nu \equiv \sqrt{n^{2} - 1}$ and 
$A,B,C$ are constants. Now we impose magnetic 
boundary conditions at the 3-sphere boundary of radius $a$. These 
set to zero at the boundary the 
tangential components of the electromagnetic potential, 
the gauge-averaging functional and hence the Faddeev-Popov ghost 
field [18, 19]. 
They lead, in the gauge (4.3), 
to Dirichlet boundary conditions for $f_n$, $g_n$ 
and ghost modes, and to Robin boundary conditions for $R_n$ modes.

On imposing the boundary conditions described above, one obtains [21]
$$
I_{\nu}(Ma)=0
\eqno (4.7)
$$
for $g_{n}$ and $\epsilon_n$, and
$$
I_{\nu}'(Ma)=0
\eqno (4.8)
$$
for $R_{n}$. Note that, since $R_1$ is proportional 
to $I_{0}(M\tau)/\tau$ in our gauge, 
the decoupled mode for normal photons has to vanish, 
to ensure regularity at 
the origin (see [21]).

First, we evaluate the contribution of $\epsilon_1$ to 
the 1-loop effective action. Since $\epsilon_{1}$ is
proportional to $I_{0}(M\tau)$, the analysis in [21],
jointly with the asymptotic expansion of $I_{0}$ at
large argument, implies (by virtue of the ghost 
degeneracy)
$$
\zeta_{\epsilon_{1}}'(0)={1\over 2}\log(a^{2})
+ \log(2\pi).
\eqno (4.9)
$$

We now study the contributions of all modes for $n \geq 2$
to the 1-loop effective action (with the exception of the
transverse modes, whose effect was evaluated in [11, 12],
as shown below). Since the degeneracy of 
$\epsilon_n$ is $-2n^{2}$, 
whereas that for $g_n$ and $R_n$ is $n^2$,  
and bearing in mind that the $R_n$ obey Robin boundary conditions 
with $u=0$ (cf [10]), one can write $\zeta(s)$ as
$$
\lim_{m \to 0}
\sum_{n=2}^{\infty}n^{2}
{\sin(\pi s) \over \pi}\int_{ma \over\nu}^{\infty} 
dz \; \Bigr[{\Bigr({z\nu \over a}\Bigr)}^{2}- m^{2}\Bigr]^{-s} \;
{\partial \over \partial z}\Bigr[ \log{z\nu \over a}
I_{\nu}'(z\nu)  - \log I_{\nu}(z\nu)\Bigr] \; .
$$
Note that $m$ is a mass parameter which differs from $M$
used so far.
Following [10], by subtracting and adding the leading terms of the 
uniform asymptotic expansions of Bessel functions and 
their first derivatives, one finds
$$
\zeta_{n \geq 2}(s)=
\lim_{m \to 0}\left(
\sum_{n=2}^{\infty}n^{2}Z_{\nu}(s)
+ {\widetilde A}_{0}(s)
+\sum_{i=1}^{3}{\widetilde A}_{i}(s)\right)
\eqno (4.10)
$$
where (see the appendix)
$$ \eqalignno{
Z_{\nu}(s)&=
{\sin(\pi s) \over \pi}\int_{ma \over\nu}^{\infty} 
dz \; \Bigr[{\Bigr({z\nu \over a}\Bigr)}^{2}- m^{2}\Bigr]^{-s} \;
{\partial \over \partial z}\Bigr\{\log\Big[{z \over (1+z^{2})^{1/2}}
{I_{\nu}'(z\nu) \over 
I_{\nu}(z\nu)}\Bigr] \cr
&+\sum_{i=1}^{3}{(D_{i}(t)-M_{i}(t,0))\over \nu^{i}} \Bigr \}
&(4.11)\cr}
$$
$$
{\widetilde A}_{0}(s)=\sum_{n=2}^{\infty}n^{2} 
{\widetilde A}_{0}^{\nu}(s)=
\sum_{n=2}^{\infty} n^{2} {\sin (\pi s) \over \pi}
\int_{ma \over \nu}^{\infty}  
dz \; \Bigr[{\Bigr({z\nu \over a}\Bigr)}^{2}- m^{2}\Bigr]^{-s} \;
{\partial \over \partial z}\ \log\Bigr(1+ z^{2}\Bigr)^{1/2} 
\eqno (4.12)
$$
$$
{\widetilde A}_{i}(s)=\sum_{n=2}^{\infty}
n^{2} {\sin(\pi s)\over \pi}
\int_{ma\over \nu}^{\infty}dz \; 
\Bigr[\Bigr({z\nu \over a}\Bigr)^{2}-m^{2}\Bigr]^{-s}
{\partial \over \partial z}
\Bigr({(M_{i}(t,0)-D_{i}(t))\over \nu^{i}}\Bigr).
\eqno (4.13)
$$
Remarkably, in this case the total 
$A_{-1}$ vanishes, since 
the terms resulting from $\exp (\nu\eta)$ cancel each other (see the 
appendix). 

On using the analyticity of $Z_{\nu}(s)$ 
in the neighbourhood of $s=0$, one finds 
the derivative $Z_{\nu}'(0)$ as (cf [10])
$$
Z_{\nu}'(0) = - \left[\log{z \over (1+z^{2})^{1/2}} {I_{\nu}'(z\nu) 
\over I_{\nu} (z\nu)}
+\sum_{i=1}^{3}{(D_{i}(t)-M_{i}(t,0))\over \nu^{i}} 
\right]_{z={ma \over \nu}} 
\eqno (4.14)
$$
and as $m \to 0$, this becomes (since $M_{i}(1,0)=D_{i}(1)$)
$$
Z_{\nu}'(0) = 0. 
\eqno (4.15)
$$
Thus, the contribution of the first infinite sum in (4.10) vanishes.

Moreover, since $u=0$ (as stated above), and 
bearing in mind the polynomials 
$M_{i}(t,0)$ and $D_{i}(t)$, 
the sum of the effects of the ${\widetilde A}_i$, for 
$i=1,2,3$, yields in the 
massless limit (see the end of the appendix)
$$
\lim_{m \to 0}
\left[{d\over ds}\sum_{i=1}^{3}
{\widetilde A}_{i}(s)\right]_{s=0}
=-{319\over 1260}.
\eqno (4.16)
$$

Last, following [10], one immediately  finds 
$$
{\widetilde A}_{0}^{\nu}(s)
={1 \over 2} m^{-2s} {}_2F_{1}\Bigr(1;s;1;
-\Bigr({\nu \over ma}\Bigr)^{2}\Bigr) = 
{1 \over 2} m^{-2s} \Bigr[1 + {({n^2} - 1) \over (ma)^{2}}\Bigr]^{-s} 
\; .
\eqno (4.17) 
$$
To evaluate the effect of ${\widetilde A}_{0}(s)$, 
it is sufficient to pick
out the coefficient of $s$ in its expansion in powers
of $s$. One then deals with the infinite sum
$$
\Sigma_{1} \equiv {1\over 2}\log(a^{2})\sum_{n=2}^{\infty}n^{2}
-{1\over 2} \lim_{y \to 0} 
\sum_{n=2}^{\infty}n^{2-2y}\log(n^{2}-1).
\eqno (4.18)
$$
Within the framework of $\zeta$-function 
regularization, the first sum in
(4.18) yields $-{1\over 2}\log(a^{2})$, whilst the finite part
of the second sum can be obtained after expressing $(n^{2}-1)$
as $(n-1)(n+1)$, and then taking the limit
$$
-{1\over 2}\lim_{y \to 0}\biggr(\sum_{n=1}^{\infty}
(n+1)^{2-2y}\log(n)+\sum_{n=3}^{\infty}(n-1)^{2-2y}\log(n)
\biggr).
$$
One thus finds
$$
\lim_{m \to 0}
\left[{d\over ds}{\widetilde A}_{0}(s)\right]_{s=0}
=\zeta_{R}'(-2)+\zeta_{R}'(0)+{1\over 2}\log(2)
-{1\over 2}\log(a^{2}).
\eqno (4.19)
$$
In the formalism of section 3, the result (4.19) reflects
the contributions of $I^{R}(0)$ and $I_{\rm reg}^{R}(\infty)$
to $\zeta'(0)$ [15--17].

Interestingly, the full $\zeta'(0)$ value differs from the 
contribution of transverse modes by the amount
$$
\zeta'(0)-\zeta_{T}'(0)= -{319\over 1260} + \log(2)
+{1\over 2}\log(\pi)+\zeta_{R}'(-2)
\eqno (4.20)
$$
since $\zeta_{R}'(0)=-{1\over 2}\log(2\pi)$.
Last, bearing in mind the
full $\zeta(0)$ value $-77/180$ [21]
and the contribution of the transverse 
modes $f_n$ to $\zeta'(0)$ [11, 12]
$$
\zeta_{\rm T}'(0)= -{6127 \over 15120} -{29 \over 45}\log(2) 
-{77 \over 90}\log(a) - \log(\pi) + {2 \over 3} \zeta_{R}'(-3) 
-\zeta_{R}'(-2)- {5\over 3}\zeta_{R}'(-1)
\eqno (4.21)
$$
and combining (4.20) and (4.21), one 
can write the  full 1-loop effective action 
for Euclidean Maxwell theory, in the gauge (4.3), as (see 1.7)
$$
\Gamma^{(1)}= {77\over 360}\log(\mu^{2}a^{2})
+{1991 \over 6048}
+{1\over 4}\log(\pi)-{8\over 45}\log(2)
-{1\over 3} \zeta_{R}'(-3) 
+{5\over 6}\zeta_{R}'(-1). 
\eqno (4.22)
$$
\vskip 0.3cm
\leftline {\bf 5. Concluding remarks}
\vskip 0.3cm
\noindent
First, our paper has compared two analytic techniques 
[10, 15--17] for the
evaluation of $\zeta'(0)$ for manifolds with boundary, showing
that they agree. In section 4, these techniques have
been applied to the calculation of the 1-loop effective action
for Euclidean Maxwell theory subject to magnetic boundary
conditions. On studying the particular gauge condition (4.3),
which leads to a complete decoupling of longitudinal, normal
and ghost modes from the beginning, such contributions have
been found to yield a non-vanishing contribution to the full
$\zeta'(0)$ value within the framework of Faddeev-Popov formalism.
This result appears interesting, since it was found in [21] that
the gauge condition (4.3) leads to a full $\zeta(0)$ value which
actually coincides with the contribution 
of the transverse modes $f_{n}$,
i.e. $\zeta(0)=\zeta_{T}(0)=-{77\over 180}$ [22]. 

In other words, the detailed calculations of section 4 add
evidence in favour of longitudinal, normal and ghost modes
playing an essential role in obtaining the correct form
of 1-loop calculations [19--21, 25]. However, the problem
remains to identify unambiguously the unphysical modes of the
quantum theory [26].
As far as we know, our paper has presented the first mode-by-mode
calculation of the 1-loop effective action for gauge fields in
the presence of boundaries, when {\it all} perturbative modes
are taken into account (the work in [11, 12] only considered the 
contribution of the so-called physical degrees of freedom, i.e. 
the transverse part of the electromagnetic potential, or
transverse-traceless perturbations for linearized gravity). 
The extension to a broader class of relativistic gauges for 
Euclidean Maxwell theory [18, 19, 21], or to higher-spin fields,
cannot be treated by hand, since it involves the uniform
asymptotic expansions of determinants of $2 \times 2$ and
$4 \times 4$ matrices. However, such a task is accessible to
modern computer programmes, and we are confident that it can be
accomplished in the near future. This is the next natural
step, since in covariant gauges the differential operators
are minimal, and one also achieves a well defined 3+1
decomposition of the background 4-geometry [19--21]. 

The corresponding {\it geometric}
form of $\zeta'(0)$ for gauge fields and gravitation in
the presence of boundaries is, however, much more difficult 
to obtain, and requires a greater effort. We thus hope that the
research described in our paper will provide the first step
towards the completion of DeWitt's effective action programme
within the framework of 1-loop quantum cosmology [27] 
and quantum gravity [28, 29].
\vskip 0.3cm
\leftline {\bf Acknowledgments}
\vskip 0.3cm
\noindent
We are much indebted to Klaus Kirsten for correspondence 
and for enlightening conversations on his work on functional
determinants. The work of A Kamenshchik was partially supported
by the Russian Foundation for Fundamental Researches 
through grant No 96-02-1000, and by the Russian Research
Project ``Cosmomicrophysics".
\vskip 10cm
\leftline {\bf Appendix}
\vskip 0.3cm
\noindent
In section 2, the equation obeyed by the eigenvalues by
virtue of the boundary conditions is
$$
I_{l+1}(\sqrt{\lambda}a)=0
\; \; \; \; {\rm for \; all} \; 
l=0,1,2,...
\eqno (A.1)
$$
where $I_{\rho}$ denotes, as usual, the regular Bessel function of
order $\rho$. The uniform asymptotic expansion of $I_{\rho}(\rho z)$
as $\rho \rightarrow \infty$ is given by [30]
$$
I_{\rho}(\rho z) \sim {1\over \sqrt{2\pi \rho}}
{e^{\rho \eta}\over (1+z^{2})^{1\over 4}}
\left[1+\sum_{k=1}^{\infty}{u_{k}(t)\over \rho^{k}}
\right]
\eqno (A.2)
$$
where $t \equiv 1/\sqrt{1+z^{2}}$,
$\eta \equiv \sqrt{1+z^{2}}+\log[z/(1+\sqrt{1+z^{2}})]$,
and the $u_{k}$ polynomials are the Debye polynomials described 
in [30]. The 1-loop analysis makes it necessary to consider
$\log(I_{\rho}(\rho z))$ [10, 15--17], and hence it is useful to 
derive the asymptotic expansion
$$
\log \left[1+\sum_{k=1}^{\infty}{u_{k}(t)\over \rho^{k}}
\right] \sim \sum_{n=1}^{\infty}{D_{n}(t)\over \rho^{n}}
\eqno (A.3)
$$
where
$$
D_{1}(t)=\sum_{j=0}^{1}x_{1,j}t^{1+2j}
={1\over 8}t-{5\over 24}t^{3}
\eqno (A.4)
$$
$$
D_{2}(t)=\sum_{j=0}^{2}x_{2,j}t^{2+2j}
={1\over 16}t^{2}-{3\over 8}t^{4}
+{5\over 16}t^{6}
\eqno (A.5)
$$
$$
D_{3}(t)=\sum_{j=0}^{3}x_{3,j}t^{3+2j}
={25\over 384}t^{3}-{531\over 640}t^{5}
+{221\over 128}t^{7}-{1105\over 1152}t^{9}.
\eqno (A.6)
$$
The identity (2.8) relies on the integral representation
of $\log \Gamma(z)$, i.e. [10, 23]
$$ 
\log \Gamma(z)=\Bigr(z-{1\over 2}\Bigr)\log(z)
-z+{1\over 2}\log(2\pi) 
+\int_{0}^{\infty}\biggr({1\over 2}-{1\over t}
+{1\over (e^{t}-1)}\biggr){e^{-tz}\over t} \; dt .
\eqno (A.7) 
$$
The contribution of (2.8) to $\zeta'(0)$ is obtained by
taking the infinite sum 
$\sum_{l=0}^{\infty}(l+1)^{2}Z_{l+1}'(0)$. A double
integration by parts, and then the use of the identity
$$
\sum_{l=0}^{\infty}e^{-lt}={1\over (1-e^{-t})}
\eqno (A.8)
$$
lead to a sum of divergent contributions which add up to
zero, plus the following term:
$$ \eqalignno{
Z'(0,z)&={1\over 360}\int_{0}^{\infty}{t^{z}e^{-t}\over
(1-e^{-t})}dt
+\int_{0}^{\infty}{t^{z-3}e^{-t}\over (1-e^{-t})}dt \cr
&-6 \int_{0}^{\infty}{t^{z-4}e^{-t}\over (1-e^{-t})}dt
+\int_{0}^{\infty}{t^{z}e^{-t}\over (1-e^{-t})}
{d^{2}\over dt^{2}}
\left({1\over t}{e^{-t}\over (1-e^{-t})}\right) dt
&(A.9)\cr}
$$
where the parameter $z$ has been introduced to regularize
the divergences of the calculation [10]. Thus, on using the
identities ($\zeta_{H}$ being the Hurwitz $\zeta$-function) 
$$
\int_{0}^{\infty}{x^{\sigma-1}e^{-\omega x}\over
(1-e^{-x})}dx=\Gamma(\sigma) \zeta_{H}(\sigma,\omega)
\eqno (A.10)
$$
$$
\int_{0}^{\infty}{x^{\sigma}e^{-(\omega+1)x}\over
(1-e^{-x})^{2}}dx=\Gamma(\sigma+1)
\Bigr[\zeta_{H}(\sigma,\omega)
-\omega \zeta_{H}(\sigma+1,\omega)\Bigr]
\eqno (A.11)
$$
$$ \eqalignno{
\int_{0}^{\infty}{x^{\sigma+1}e^{-(\omega+2)x}\over
(1-e^{-x})^{3}}dx
&={1\over 2}\Gamma(\sigma+2)
\Bigr[\zeta_{H}(\sigma,\omega)-(1+2\omega)
\zeta_{H}(\sigma+1,\omega)\cr
&+\omega(1+\omega)\zeta_{H}(\sigma+2,\omega)\Bigr]
&(A.12)\cr}
$$
$$ \eqalignno{
\; & \int_{0}^{\infty}{x^{\sigma+2}e^{-(\omega+3)x}\over
(1-e^{-x})^{4}}dx
={1\over 6}\Gamma(\sigma+3)\Bigr[\zeta_{H}(\sigma,\omega)
-3(1+\omega)\zeta_{H}(\sigma+1,\omega) \cr
&+(2+6\omega+3\omega^{2})\zeta_{H}(\sigma+2,\omega)
-\omega(2+3\omega+\omega^{2})\zeta_{H}(\sigma+3,\omega)
\Bigr]
&(A.13)\cr}
$$
one finds
$$ \eqalignno{
Z'(0,z)&={1\over 360}\Gamma(z+1)\zeta_{R}(z+1)
+\Bigr[{1\over 3}z^{3}-z^{2}+{2\over 3}z-8 \Bigr]\Gamma(z-3)
\zeta_{R}(z-3) \cr
&+\Bigr[-{1\over 2}z^{2}+{1\over 2}z \Bigr]
\Gamma(z-2)\zeta_{R}(z-2) 
+{1\over 6}(z-1)\Gamma(z-1)\zeta_{R}(z-1).
&(A.14)\cr}
$$
At this stage, one can insert into (A.14) and (2.3)--(2.7)
the expansions (as $\varepsilon \rightarrow 0$)
$$
\Gamma(\varepsilon-3)=-{1\over 6}{1\over \varepsilon}
+ {1 \over 6}\gamma -{11\over 36} 
+{\rm O}(\varepsilon)
\eqno (A.15)
$$
$$
\Gamma(\varepsilon-2)={1\over 2}{1\over \varepsilon}
-{1\over 2}\gamma +{3\over 4}+{\rm O}(\varepsilon)
\eqno (A.16)
$$
$$
\Gamma(\varepsilon-1)=-{1\over \varepsilon}+\gamma-1
+{\rm O}(\varepsilon)
\eqno (A.17)
$$
$$
\Gamma(\varepsilon)={1\over \varepsilon}-\gamma
+{\rm O}(\varepsilon)
\eqno (A.18)
$$
$$
\Gamma \Bigr(\varepsilon-{1\over 2}\Bigr)=-2\sqrt{\pi}
\Bigr[1+\varepsilon(-\gamma-2\log(2)+2)
+{\rm O}(\varepsilon^{2})\Bigr]
\eqno (A.19)
$$
$$
\zeta_{R}(1+\varepsilon)={1\over \varepsilon}+\gamma
+{\rm O}(\varepsilon)
\eqno (A.20)
$$
$$
\zeta_{R}(-n+\varepsilon)=\zeta_{R}(-n)
+\varepsilon \; \zeta_{R}'(-n)
+{\rm O}(\varepsilon^{2})
\eqno (A.21)
$$
and the identities [30]
$$
\zeta_{H}(s,1)=\zeta_{R}(s)
\eqno (A.22)
$$
$$
\zeta_{H} \Bigr(s,{1\over 2}\Bigr)=(2^{s}-1)\zeta_{R}(s).
\eqno (A.23)
$$
This leads to the result (2.9) for Dirichlet boundary conditions.

For Robin boundary conditions, we refer the reader to section 4
of [10], bearing in mind that the polynomials (A.4)--(A.6) are
replaced by new polynomials $M_{n}(t,u)$ given by [10, 31]
$$
M_{1}(t,u)=\Bigr(-{3\over 8}+u \Bigr)t+{7\over 24}t^{3}
\eqno (A.24)
$$
$$
M_{2}(t,u)=\Bigr(-{3\over 16}+{1\over 2}u-{1\over 2}u^{2}\Bigr)t^{2}
+\Bigr({5\over 8}-{1\over 2}u\Bigr)t^{4}-{7\over 16}t^{6}
\eqno (A.25)
$$
$$ \eqalignno{
M_{3}(t,u)&=\Bigr(-{21\over 128}+{3\over 8}u-{1\over 2}u^{2}
+{1\over 3}u^{3}\Bigr)t^{3}
+\Bigr({869\over 640}-{5\over 4}u+{1\over 2}u^{2}\Bigr)t^{5}\cr
&+\Bigr(-{315\over 128}+{7\over 8}u \Bigr)t^{7}
+{1463\over 1152}t^{9}
&(A.26)\cr}
$$
where $u$ is the dimensionless parameter occurring in the formula
for Robin boundary conditions [10]. In our section 4,
$u$ vanishes.

In section 3, the integral in (3.4) is evaluated by means of
the general formula
$$
\int_{0}^{\infty}{\log(x)\over (x+1)^{m+1/2}}dx
={2\over (2m-1)}\left(2\log(2)-\sum_{k=1}^{m-2}
{1\over k}-2\sum_{k=m-1}^{2m-3}{1\over k} \right)
\eqno (A.27)
$$
which leads, in particular, to
$$
\int_{0}^{\infty}{\log(x)\over (x+1)^{5/2}}dx
={4\over 3}\log(2)-{4\over 3}
\eqno (A.28)
$$
$$
\int_{0}^{\infty}{\log(x)\over (x+1)^{7/2}}dx
={4\over 5}\log(2)-{16\over 15}
\eqno (A.29)
$$
$$
\int_{0}^{\infty}{\log(x)\over (x+1)^{9/2}}dx
={4\over 7}\log(2)-{92\over 105}
\eqno (A.30)
$$
$$
\int_{0}^{\infty}{\log(x)\over (x+1)^{11/2}}dx
={4\over 9}\log(2)-{704\over 945}.
\eqno (A.31)
$$

In section 4, the $Z_{\nu}(s)$ and 
the ${\widetilde A}_{i}(s)$, 
for $i = -1,\ldots,3 $ are the sum of the contributions
resulting from $g_n$, $R_n$ and 
$\epsilon_n$ with their own degeneracies.
For the $g_n$ and $\epsilon_n$ modes
(obeying Dirichlet boundary conditions) one has [10, 32]
$$\eqalignno{
Z_{\nu}^{D}(s)&= {\sin (\pi s) \over \pi} 
\int_{ma \over \nu}^{\infty} \;
dz\; {\Bigr[{\Bigr({z\nu \over a}\Bigr)}^{2} - m^{2}\Bigr]}^{-s} 
{\partial \over \partial z} \Bigr\{ \log I_{\nu}(z\nu) \cr
&- \log \Bigr[{1 \over \sqrt{2\pi\nu}} 
{e^{\nu\eta} \over (1+z^{2})^{1 \over 4}}\Bigr] 
- \sum_{k=1}^{3}{D_{k}(t) \over \nu^{k}}\Bigr\}
&(A.32)\cr}
$$
$$
A_{-1}^{\nu,D} ={\sin (\pi s) \over \pi} 
\int_{ma \over \nu}^{\infty} \;
dz \; {\Bigr[{\Bigr({z\nu \over a}\Bigr)}^{2} - m^{2}\Bigr]}^{-s}
{\partial \over \partial z} 
\log \Bigr({z^{-\nu} \over \sqrt{2\pi\nu}} e^{\nu\eta}\Bigr)
\eqno (A.33)
$$
$$
A_{0}^{\nu,D} ={\sin (\pi s) \over \pi} \int_{ma \over \nu}^{\infty} \;
dz\; \Bigr[\Bigr({z\nu \over a}\Bigr)^{2} - m^{2}\Bigr]^{-s}
{\partial \over \partial z} \log(1+z^{2})^{- {1 \over 4}}
\eqno (A.34)
$$
$$
A_{i}^{\nu,D} ={\sin (\pi s) \over \pi} \int_{ma \over \nu}^{\infty} \;
dz \; \Bigr[\Bigr({z\nu \over a}\Bigr)^{2} - m^{2}\Bigr]^{-s}
{\partial \over \partial z}\Bigr({D_{i}(t) \over \nu^{i}}\Bigr)\; .
\eqno (A.35)
$$

For the $R_{n}$ modes, 
which obey Robin boundary conditions with $u=0$, one has
$$\eqalignno{
Z_{\nu}^{R}(s)&= {\sin (\pi s) \over \pi} \int_{ma \over \nu}^{\infty} \;
dz\; \Bigr[\Bigr({z\nu \over a}\Bigr)^{2} - m^{2}\Bigr]^{-s} 
{\partial \over \partial z} 
\Bigr\{\log\Bigr[{z\nu \over a} I_{\nu}'(z\nu)\Bigr]\cr
&- \log \Bigr[{\nu \over a\sqrt{2\pi\nu}} e^{\nu\eta} 
(1+z^{2})^{1 \over 4}\Bigr] 
- \sum_{k=1}^{3}{M_{k}(t,0) \over \nu^{k}}\Bigr\}
&(A.36)\cr}
$$
$$
A_{-1}^{\nu,R}=A_{-1}^{\nu,D}
\eqno (A.37)
$$
$$
A_{0}^{\nu,R}=-A_{0}^{\nu,D}
\eqno (A.38)
$$
and 
$$
A_{i}^{\nu,R} ={\sin (\pi s) \over \pi} \int_{ma \over \nu}^{\infty} \;
dz \; 
\Bigr[\Bigr({z\nu \over a}\Bigr)^{2} - m^{2}\Bigr]^{-s}
{\partial \over \partial z}
\Bigr({M_{i}(t,0) \over \nu^{i}}\Bigr) \; .
\eqno (A.39)
$$
The result (4.16) holds by virtue of the identities [32]
$$ \eqalignno{
\; & \int_{ma\over \nu}^{\infty} \; dz 
\Bigr[\Bigr({z\nu \over a}\Bigr)^{2} - m^{2}\Bigr]^{-s}
{\partial \over \partial z}t^{l} \cr
&=-{1\over 2}m^{-2s}{l\over (ma)^{l}}
{\Gamma \Bigr(s+{1\over 2}l\Bigr)\Gamma(1-s)\over
\Gamma \Bigr(1+{1\over 2}l\Bigr)}\nu^{l}
\Bigr[1+\Bigr({\nu \over ma}\Bigr)^{2}\Bigr]^{-s-{l\over 2}}
&(A.40)\cr}
$$
$$
{\sin(\pi s)\over \pi}\Gamma(s)\Gamma(1-s)=1
\eqno (A.41)
$$
jointly with the formulae
$$
M_{1}(t,0)-D_{1}(t)=-{1\over 2}t+{1\over 2}t^{3}
\eqno (A.42)
$$
$$
M_{2}(t,0)-D_{2}(t)=-{1\over 4}t^{2}+t^{4}
-{3\over 4}t^{6}
\eqno (A.43)
$$
$$
M_{3}(t,0)-D_{3}(t)=-{11\over 48}t^{3}+{35\over 16}t^{5}
-{67\over 16}t^{7}+{107\over 48}t^{9}
\eqno (A.44)
$$
which result from (A.4)--(A.6) and (A.24)--(A.26). 
More precisely, denoting by ${\tilde x}_{l,j}$ the numerical
coefficients occurring in (A.42)--(A.44), the insertion of
(A.40) into (4.13) yields (cf (2.5)--(2.7))
$$
{\widetilde A}_{1}(s)=-{1\over 2}{a^{2s}\over \Gamma(s)}
\sum_{j=0}^{1}{\tilde x}_{1,j}(1+2j)
{\Gamma(s+j+{1\over 2})\over \Gamma(j+{3\over 2})}
\sum_{n=2}^{\infty}n^{2}(n^{2}-1)^{-s-{1\over 2}}
\eqno (A.45)
$$
$$
{\widetilde A}_{2}(s)=-{1\over 2}{a^{2s}\over \Gamma(s)}
\sum_{j=0}^{2}{\tilde x}_{2,j}(2+2j)
{\Gamma(s+j+1)\over \Gamma(j+2)}
\sum_{n=2}^{\infty}n^{2}(n^{2}-1)^{-s-1}
\eqno (A.46)
$$
$$
{\widetilde A}_{3}(s)=-{1\over 2}{a^{2s}\over \Gamma(s)}
\sum_{j=0}^{3}{\tilde x}_{3,j}(3+2j)
{\Gamma(s+j+{3\over 2})\over \Gamma(j+{5\over 2})}
\sum_{n=2}^{\infty}n^{2}(n^{2}-1)^{-s-{3\over 2}}.
\eqno (A.47)
$$
These formulae make it necessary to use the expansion [17]
$$
(n^{2}-1)^{-s}=\sum_{r=0}^{\infty}
{\Gamma(r+s)\over \Gamma(s)\Gamma(r+1)}n^{-2r-2s}.
\eqno (A.48)
$$
One can then insert (A.48) into (A.45)--(A.47), bearing in
mind the expansions (A.18)--(A.20) and the property
$\sum_{j=0}^{l}{\tilde x}_{l,j}=0$, for all $l=1,2,3$, 
which holds by virtue of (A.42)--(A.44). This implies that
${\widetilde A}_{1}(s)$ contributes $-{1\over 4}$ to
$\zeta'(0)$, which results from the expansion of 
$\zeta_{R}(2s+1)$ in the identity
$$ \eqalignno{
\sum_{n=2}^{\infty}n^{2}(n^{2}-1)^{-s-{1\over 2}}
&=\zeta_{R}(2s-1)-1+\Bigr(s+{1\over 2}\Bigr)
\Bigr(\zeta_{R}(2s+1)-1 \Bigr)\cr
&+\sum_{r=2}^{\infty}{\Gamma(r+s+{1\over 2})\over
\Gamma (s+{1\over 2})\Gamma(r+1)}
\Bigr[\zeta_{R}(2r+2s-1)-1 \Bigr]
&(A.49)\cr}
$$
after insertion into (A.45). Moreover, ${\widetilde A}_{2}(s)$
yields a vanishing contribution, whilst ${\widetilde A}_{3}(s)$
contributes
$$
-{1\over 6}\Bigr(2{\tilde x}_{3,1}+{16\over 5}{\tilde x}_{3,2}
+{142\over 35}{\tilde x}_{3,3}\Bigr)=-{1\over 315}
$$
since $\sum_{n=2}^{\infty}n^{2}(n^{2}-1)^{-s-{3\over 2}}
=\zeta_{R}(2s+1)-1+\sum_{r=1}^{\infty}
{\Gamma(r+s+{3\over 2})\over \Gamma(s+{3\over 2})\Gamma(r+1)}
\Bigr[\zeta_{R}(2r+2s+1)-1 \Bigr]$, by virtue of (A.48).
Hence one finds the value given in (4.16).

In the formalism of section 3, the same result is obtained 
by evaluating integrals of the kind
$$
J_{i} \equiv
\int_{0}^{\infty}dz \; \log(z) {\partial \over \partial z^{2}}
\Bigr(M_{i}(t,0)-D_{i}(t)\Bigr)
\eqno (A.50)
$$
and then using (A.27)--(A.31). For example, (A.27) and (A.28)
imply that $J_{1}=1$. This contribution should be divided by
$2$, bearing in mind that, as $n \rightarrow \infty$,
${1\over \nu} \sim {1\over n}\biggr(1+{1\over 2}{1\over n^{2}}
+{\rm O}(n^{-4})\biggr)$. Moreover, $J_{2}$ does not contribute to
$\zeta'(0)$, whilst $J_{3}={2\over 315}$, and the result (4.16)
is obtained, according to [15--17], as
$-{1\over 2}\Bigr({1\over 2}+{2\over 315}\Bigr)$.
\vskip 0.3cm
\leftline {\bf References}
\vskip 0.3cm
\item {[1]}
DeWitt B S 1965 {\it Dynamical Theory of Groups and Fields}
(New York: Gordon and Breach)
\item {[2]}
DeWitt B S 1975 {\it Phys. Rep.} {\bf 19} 295
\item {[3]}
DeWitt B S 1984 in {\it Relativity, Groups and Topology II}
eds B S DeWitt and R Stora (Amsterdam: North Holland)
\item {[4]}
Vilkovisky G A 1984 {\it Nucl. Phys.} B {\bf 234} 125
\item {[5]}
DeWitt B S 1987 in {\it The Architecture of Fundamental
Interactions at Short Distances} eds P Ramond and R Stora 
(Amsterdam: North-Holland)
\item {[6]}
Buchbinder I L, Odintsov S D and Shapiro I L 1992 
{\it Effective Action in Quantum Gravity} 
(Bristol: IOP Publishing)
\item {[7]}
Avramidi I G 1991 {\it Nucl. Phys.} B {\bf 355} 712
\item {[8]}
Barvinsky A O, Gusev Yu V, Vilkovisky G A and Zhytnikov
V V 1995 {\it Nucl. Phys.} B {\bf 439} 561
\item {[9]}
Dowker J S and Apps J S 1995 {\it Class. Quantum Grav.}
{\bf 12} 1363
\item {[10]}
Bordag M, Geyer B, Kirsten K and Elizalde E 1995 
Zeta-function determinant of the Laplace operator
on the D-dimensional ball 
{\it HEP-TH} 9505157 to appear in {\it Comm. Math. Phys.}
\item {[11]}
Dowker J S 1995 Spin on the 4-ball {\it HEP-TH} 9508082
\item {[12]}
Kirsten K and Cognola G 1995 Heat-kernel coefficients
and functional determinants for higher-spin fields
on the ball {\it HEP-TH} 9508088
\item {[13]}
Schwinger J 1951 {\it Phys. Rev.} {\bf 82} 664
\item {[14]}
Seeley R T 1967 {\it Amer. Math. Soc. Proc. Symp.
Pure Math.} {\bf 10} 288
\item {[15]}
Barvinsky A O, Kamenshchik A Yu, Karmazin I P and
Mishakov I V 1992 {\it Class. Quantum Grav.} {\bf 9} L27
\item {[16]}
Barvinsky A O, Kamenshchik A Yu and Karmazin I P 1992
{\it Ann. Phys.} (N.Y.) {\bf 219} 201
\item {[17]}
Kamenshchik A Yu and Mishakov I V 1992 
{\it Int. J. Mod. Phys.} A {\bf 7} 3713
\item {[18]}
Esposito G 1994 {\it Class. Quantum Grav.} 
{\bf 11} 905
\item {[19]}
Esposito G, Kamenshchik A Yu, Mishakov I V and Pollifrone G
1994 {\it Class. Quantum Grav.} {\bf 11} 2939
\item {[20]}
Esposito G and Kamenshchik A Yu 1994 {\it Phys. Lett.} 
{\bf 336B} 324
\item {[21]}
Esposito G, Kamenshchik A Yu, Mishakov I V and Pollifrone G
1995 {\it Phys. Rev.} D {\bf 52} 2183
\item {[22]}
Louko J 1988 {\it Phys. Rev.} D {\bf 38} 478
\item {[23]}
Bateman H and Erdelyi A 1953 {\it Higher Transcendental
Functions} Vol 1 (New York: McGraw-Hill)
\item {[24]}
Lifshitz E M and Khalatnikov I M 1963 {\it Adv. Phys.}
{\bf 12} 185
\item {[25]}
Moss I G and Poletti S 1994 {\it Phys. Lett.} {\bf 333B} 326
\item {[26]}
Vassilevich D V 1995 {\it Phys. Rev.} D {\bf 52} 999
\item {[27]}
Esposito G 1994 {\it Quantum Gravity, Quantum Cosmology
and Lorentzian Geometries} 
({\it Lecture Notes in Physics} {\bf m12}) (Berlin: Springer)
\item {[28]}
Barvinsky A O, Gusev Yu V, Zhytnikov V V and Vilkovisky
G A 1995 {\it Class. Quantum Grav.} {\bf 12} 2157
\item {[29]}
Mirzabekian A G and Vilkovisky G A 1995 {\it Class. Quantum
Grav.} {\bf 12} 2173
\item {[30]}
Abramowitz M and Stegun I A 1964 {\it Handbook of Mathematical
Functions with Formulas, Graphs and Mathematical Tables}
(New York: Dover)
\item {[31]}
Moss I G 1989 {\it Class. Quantum Grav.} {\bf 6} 759
\item {[32]}
Bordag M, Elizalde E and Kirsten K 1995 Heat-kernel 
coefficients of the Laplace operator on the D-dimensional ball
{\it HEP-TH} 9503023 to appear in {\it J. Math. Phys.}

\bye